# Gibbs Sampling for (Coupled) Infinite Mixture Models in the Stick Breaking Representation

**Ian Porteous, Alex Ihler, Padhraic Smyth, Max Welling**
Department of Computer Science
UC Irvine, Irvine CA 92697-3425
{*iporteou,ihler,smyth,welling*}@ics.uci.edu

## Abstract

Nonparametric Bayesian approaches to clustering, information retrieval, language modeling and object recognition have recently shown great promise as a new paradigm for unsupervised data analysis. Most contributions have focused on the Dirichlet process mixture models or extensions thereof for which efficient Gibbs samplers exist. In this paper we explore Gibbs samplers for infinite complexity mixture models in the stick breaking representation. The advantage of this representation is improved modeling flexibility. For instance, one can design the prior distribution over cluster sizes or couple multiple infinite mixture models (e.g., over time) at the level of their parameters (i.e., the dependent Dirichlet process model). However, Gibbs samplers for infinite mixture models (as recently introduced in the statistics literature) seem to mix poorly over cluster labels. Among others issues, this can have the adverse effect that labels for the same cluster in coupled mixture models are mixed up. We introduce additional moves in these samplers to improve mixing over cluster labels and to bring clusters into correspondence. An application to modeling of storm trajectories is used to illustrate these ideas.

## 1 INTRODUCTION

Nonparametric Bayesian statistics has provided elegant solutions to certain machine learning problems, in particular in clustering, text modeling and object recognition. Importantly, Bayesian nonparametric techniques can handle questions of model selection or averaging in a principled manner. Where traditional Bayesian approaches often assume an unknown but finite number components in a model, nonparametric Bayesian approaches put their prior on infinitely complex models, resulting in a number of components that grows with the number of datacases. Moreover, with the increasing computational power of modern day computers these methods have now become feasible alternatives in many machine learning domains.

The Dirichlet process (DP) (Ferguson, 1973; Blackwell & MacQueen, 1973) is one of the most popular tools from the nonparametric Bayesian toolbox. Its popularity can be partially explained by the fact that there exist simple and intuitive Gibbs sampling procedures that work well in practice (Escobar & West, 1995; Bush & MacEachern, 1996; MacEachern & Müller, 1998; Neal, 2000; Green & Richardson, 2001). One very attractive property of these Gibbs samplers is that they sample in the space of equivalence classes over cluster identities, i.e., the probability distribution is invariant w.r.t. permutations of the class labels. This is clearly the appropriate space in which to operate, since class labels are unidentifiable and irrelevant. However, this convenient property comes with a price, namely that the distribution over cluster sizes is only partially at one's control. For a DP this distribution is exponential with decay parameter $\alpha$; for the Pitman-Yor (PYP) process power-law distributions can also be modeled.

A more general representation exists that subsumes the DP and the PYP, known as the stick-breaking representation (Sethuraman, 1994). In this representation, very general distributions over cluster sizes can be modeled. Recently several collapsed Gibbs samplers were proposed which operate in this representation (Ishwaran & James, 2001; Ishwaran & James, 2003; Papaspiliopoulos & Roberts, 2005). However, such sampling occurs in a very different space than the DP samplers mentioned above, namely in the space of cluster labels directly (rather than the space of equivalence classes). In effect, the size-biased ordering of the expected prior mixture probabilities means that a permutation of the cluster labels changes the probability distribution. A point that seems to have been missed in the literature is that the collapsed Gibbs samplers described in Ishwaran and James (2003); Papaspiliopoulos and Roberts (2005) can easily become stuck in local modes corresponding to one particular assignment of cluster la-bels to clusters. In fact, this mode is rarely the mode with highest probability. Although the favored clustering associated with any of these modes is typically very similar, we will argue that any sampler should in fact mix over these modes to remove a clustering bias. Furthermore, we will show that the addition of some simple Metropolis-Hastings moves corrects this issue.

A much more dramatic impact of this poor mixing over labels can be observed when we couple infinite mixture models together at the level of the parameters as in the dependent Dirichlet process (DDPs) (MacEachern, 2000). In this case, the poor mixing over labels has the effect that corresponding clusters are not assigned to the same cluster label. Although the probability of such an assignment is very low, the Gibbs sampler (MacEachern, 2000) is not able to mix away from it. This issue has not been noted in the literature mainly because the DDP model that is used in practice represents a special case (the single-p model) that does not seem to suffer from this problem (Gelfand et al., 2004). We propose extra moves for the Gibbs sampler to deal with this issue. An application of the resulting Gibbs sampler to tracking of storms illustrates these ideas.

## 2 GIBBS SAMPLING FOR DP MIXTURE MODELS

One way to think of DP mixture models is as the limit of a finite mixture model (of, say, Gaussian components) with a uniform Dirichlet prior. Let us denote the parameters (mean and covariance) of $K$ clusters by $\boldsymbol{\theta} = \{\boldsymbol{\mu}_i, \Sigma_i\}$ for $i = 1\ldots K$, the matrix of observed data by $X = \mathbf{x}_1, .., \mathbf{x}_N$, and let $Z = z_1, .., z_N$ be the assignment variables. For a uniform Dirichlet prior with hyperparameters $\alpha/K$, one may integrate out the Dirichlet prior (Ishwaran & Zarapour, 2002; Neal, 2000) to obtain the joint distribution $P(X, Z) = P(X|Z)P(Z)$ where

$$P(X|Z) = \int d\boldsymbol{\theta}\ P(X|Z, \boldsymbol{\theta})\ P(\boldsymbol{\theta}) \quad (1)$$

$$P(Z) = P(z_1) \prod_{n=2}^{N} P(z_n|z_1, .., z_{n-1}) \quad (2)$$

with

$$P(z_n = i|z_1, .., z_{n-1}) = \frac{\alpha/K + N_i}{\alpha + n - 1}$$

where $N_i$ is the number of data previously assigned to cluster $i$. Taking the limit as $K \to \infty$, one can express this as

$$P(z_n = i|z_1, .., z_{n-1}) = \begin{cases} \frac{N_i}{\alpha+n-1} & N_i > 0 \\ \frac{\alpha}{\alpha+n-1} & i = 1 + \max_{j<n} z_j \end{cases} \quad (3)$$

We call the conditional probability over cluster assignments $P(z_n = i|z_1, .., z_{n-1})$ a *prediction rule* and note that it is exchangeable: the total probability $P(Z)$ does not depend on the order in which we process the data. Also, note that in Eqn. (3) we have pooled together all of the (infinitely many) remaining clusters, i.e., those clusters which do not yet have any data assignments.

For the purposes of clustering, we are interested in drawing samples from the posterior $P(Z|X)$. Following the statistics literature (Escobar & West, 1995; Bush & MacEachern, 1996; MacEachern & Müller, 1998; Neal, 2000), we use a collapsed Gibbs sampler which samples an assignment for one datum $x_n$ at a time from the conditional distribution $P(z_n|Z_{(-n)}, X)$, where $Z_{(-n)}$ denotes the assignments of all data except $x_n$. The expression for this conditional follows directly from Eqns. (1)–(3) and is given by,

$$P(z_n = i|Z_{(-n)}, X) \propto P(z_n = i|Z_{(-n)}) \times$$
$$\int d\boldsymbol{\theta}_i\ P(\mathbf{x}_n|z_n = i, \boldsymbol{\theta}_i) P(\boldsymbol{\theta}_i|X_{(-n)}, Z_{(-n)}) \quad (4)$$

where

$$P(\boldsymbol{\theta}_i|X_{(-n)}, Z_{(-n)}) \propto \prod_{j|z_j=i} \mathcal{N}[\mathbf{x}_j; \boldsymbol{\mu}_i, \Sigma_i]\ P(\Sigma_i^{-1}, \boldsymbol{\mu}_i)$$

If we use a normal-Wishart prior the predictive distribution can be computed analytically as a student-t distribution (Gelman et al., 2004, p.88). The expression for $P(z_n = i|Z_{(-n)})$ is analogous to Eqn. (3), replacing $N_i$ by $N_i^{(-n)}$, the total number of data assigned to cluster $i$ if we remove data-case $n$ from the pool, and $n$ by $N$, the total number of data. This expression is easily derived using exchangeability of the data, since we may always assume that the datum under consideration is the last one in Eqn. (3).

The Gibbs sampler now simply rotates through reassigning data to clusters using the conditional distribution, Eqn. (4). Samplers using additional split and merge moves are defined in Neal (2000) and Green and Richardson (2001).

## 3 SAMPLING FROM STICK–BREAKING PRIORS

The Gibbs sampler described in the previous section depends on the prediction rule given by Eqn. (3), that is, the conditional probability of assigning a particle $x_n$ to some existing cluster or to a new cluster. The main requirement for such a prediction rule is that it defines an *exchangeable* random partition of the integers $1..N$. This implies that the joint probability of an assignment $z_1, ..., z_N$ as computed by the product in Eqn. (2) is independent of the order in which we process the data. This is in fact a very strong requirement (Pitman, 2002), and to the best of our knowledge exchangeable prediction rules over assignment partitions are only known for the DP and the Pitman-Yor process (PYP).

## 3.1 THE STICK–BREAKING REPRESENTATION

There is a more general representation that encompasses both DPs and PYPs. The "stick-breaking" representation (Sethuraman, 1994) states that the probability measure underlying a DP can be written as,

$$\mathcal{P}(\boldsymbol{\theta}|\boldsymbol{\pi}, \boldsymbol{\theta}^*) = \sum_{i=1}^{\infty} \pi_i \delta(\theta_i - \theta_i^*)$$

where $\boldsymbol{\theta}^*$ are sampled IID from the prior $H(\boldsymbol{\theta})$ and the relative cluster sizes $\boldsymbol{\pi}$ follow the stick-breaking construction,

$$\pi_i = \pi_i(V) = V_i \prod_{j=1}^{i-1}(1 - V_j)$$

where the $V_i$ are sampled IID from the beta distribution $\mathcal{B}(V\,;\,1,\alpha)$. This process can be thought of as repeatedly breaking a stick of unit length into two pieces where the breakpoint is randomly sampled from the Beta distribution. We equate $\pi_1$ to the length of the left-hand segment of the stick. We then take the right segment and break it again randomly; the left segment of this remainder is then equated to $\pi_2$, and so forth. This process guarantees that the infinite sum of mixture weights $\sum_i \pi_i$ converges to 1 (the total length of the stick) in probability.

Processes much more general than the DP can be obtained using this construction by choosing the parameters of the Beta distribution arbitrarily and making them dependent on the label $i$. In other words, after $i - 1$ breaks, we randomly break the remainder at a length drawn from $\mathcal{B}(a_i, b_i)$, and follow the same construction[1] to obtain the weights $\pi_i$. As an example, the PYP is obtained by choosing $a_i = 1 - a$ and $b_i = b + i \times a$ for $a \in [0, 1)$ and $b > -a$.

The distribution of the cluster weights is determined by the setting of the $\{a_i, b_i\}$. For instance for the DP with $a_i = 1, b_i = \alpha$ we obtain a exponential distribution of cluster sizes and for a special case of the PYP with $a_i = 1 - \beta, b_i = i \times \beta$ we find a power law distribution of the cluster sizes. The stick breaking representation thus offers a flexible means of *designing* these cluster size distributions. In particular, when we have prior information about this distribution this framework can offer a way of expressing that information.

Given $N$ particles assigned to clusters $z_1, .., z_N$, the posterior probability of mixture weights given these assignments can be computed in closed form. For a finite mixture model of $K$ clusters, it is easy to show that the joint probability is

$$P(V|Z) \propto \prod_i \mathcal{B}(V_i\,;\,a_i^*, b_i^*) \qquad (5)$$

---
[1]To make sure that the weights sum to 1 in this more general setting we must check that $\sum_{i=1}^{\infty} \mathbb{E}[\log(1 - V_i)] = -\infty$, or alternatively, it is also sufficient to check that $\sum_{i=1}^{\infty} \log[1 + a_i/b_i] = \infty$ (Ishwaran & James, 2001).

and posterior expected cluster weights[2] are given by

$$\mathbb{E}[\pi_i|Z] = \frac{a_i^*}{a_i^* + b_i^*} \prod_{j=1}^{i-1} \frac{b_j^*}{a_j^* + b_j^*} \qquad (6)$$

where $a_i^* = a_i + N_i$, $b_i^* = b_i + \sum_{j=i+1}^{\infty} N_j$ and $N_i$ is the total number of particles in cluster $i$. The proof for the same result in the infinite case can be found in Ishwaran and James (2003).

If we choose to set $a_i = \gamma_i$ and $b_i = \sum_{j=i+1}^{\infty} \gamma_i$, then $a_i + b_i = b_{i-1}$, resulting in the simplified expression

$$\mathbb{E}[\pi_i|Z] = \frac{\gamma_i + N_i}{\gamma + N} \qquad \gamma = \sum_{i=1}^{\infty} \gamma_i \qquad (7)$$

In this single parameter model (Ishwaran & James, 2003), the $\gamma_i$ play the role of pseudo-particles, i.e., a (possibly fractional) number of data assigned a priori to cluster $i$.

However, there is an important subtlety to the stick–breaking formulation which has the potential to cause some confusion. Applying the values $\{a_i = 1, b_i = \alpha\}$ to Eqn. (5), we do *not* directly obtain the prediction rule for a DP given in Eqn. (3). The difference is that the prediction rule given by Eqn. (3) operates in the space of *equivalence classes* over cluster labels (i.e., $z_1 = z_2 \neq z_3 = z_4 = z_5 \ldots$, and the actual cluster identities are irrelevant), while the stick-breaking representation is defined in the space of the explicit cluster labels (i.e., $z_1 = 2, z_2 = 2, z_3 = 8, z_4 = 8, z_5 = 8, \ldots$). Hence, the DP sampler ignores the actual labels and considers all instances where $z_1 = z_2 \neq z_3 = z_4 = z_5...$ as a single equivalence class. To drive this point home, consider sampling two data assignments from the same cluster. For a DP we know from the prediction rule that the total probability of this event is $1/(1 + \alpha)$. However, in the stick breaking representation we must sum over all labels to obtain the same result:

$$P(z_1 = z_2) = \sum_{i=1}^{\infty} P(z_2 = i|z_1 = i)P(z_1 = i)$$

which after some algebra can be found to be equal to $1/(1 + \alpha)$ as well. Clearly, the space of equivalence classes is to be preferred over the space of explicit cluster labels because nothing observable depends on these labels. At the same time, it is not easy to see how one could avoid it in the stick-breaking representation, because not all prior mixture probabilities can be equal and finite (an infinite number of them must sum to 1). The important consequence of these observations is that the sampler needs to *mix over cluster labels*. If it does not, clusters with lower labels have an unfair advantage over clusters with higher labels because they have higher prior probabilities. As it turns out, the standard Gibbs sampler described in Ishwaran and James (2003) mixes very poorly over cluster labels. We address this issue more fully in Section 3.3.

---
[2]The prior expected cluster weights are easily obtained by evaluating Eqn. (6) with the $N_i = 0$.

## 3.2 MONTE-CARLO SAMPLING

We now turn our attention to the infinite mixture model described in Ishwaran and James (2003). One of the goals of this paper is to test this algorithm as a general clustering tool and to suggest improvements where necessary.

The joint distribution of a sample $X = \mathbf{x}_1,..,\mathbf{x}_N$, parameters $\boldsymbol{\theta}_i$, $i = 1,..,\infty$, assignment variables $Z = z_1,..,z_N$ and stick-lengths $V_i$, $i = 1,..,\infty$ is given as $P(X, Z, V, \boldsymbol{\theta}) = P(X|Z, \boldsymbol{\theta})P(Z|V)P(V)P(\boldsymbol{\theta})$, where

$$P(X|Z,\boldsymbol{\theta}) = \prod_{n=1}^{N} \mathcal{N}[\mathbf{x}_n; \boldsymbol{\mu}_{z_n}, \Sigma_{z_n}] \tag{8a}$$

$$P(Z|V) = \prod_{n=1}^{N} \pi_{z_n}(V) \tag{8b}$$

$$P(V) = \prod_{i=1}^{\infty} \mathcal{B}(V_i\,;\,a_i, b_i) \tag{8c}$$

$$P(\boldsymbol{\theta}) = \prod_{i=1}^{\infty} \mathcal{NW}[\boldsymbol{\mu}_i, \Sigma_i^{-1}] \tag{8d}$$

with $\mathcal{NW}$ the normal-Wishart prior for cluster means and inverse covariances.

The collapsed Gibbs sampler iteratively samples from the conditional distribution $P(z_n = i|Z_{(-n)}, X)$ computed as,

$$\begin{aligned}
&P(z_n = i|Z_{(-n)}, X) \propto \\
&\int d\boldsymbol{\theta}_i\; P(\mathbf{x}_n|z_n = i, \boldsymbol{\theta}_i)P(\boldsymbol{\theta}_i|X_{(-n)}, Z_{(-n)} = i)\times \\
&\int dV\; P(z_n = i|V)P(V|Z_{(-n)}) \qquad i = 1..\infty
\end{aligned} \tag{9}$$

where $P(\boldsymbol{\theta}_i|X_{(-n)}, Z_{(-n)})$ and $P(V|Z_{(-n)})$ are the posterior distribution of the parameters and the stick-lengths respectively.

If we use the Normal-Wishart conjugate prior, the predictive distribution $P(\mathbf{x}_n|Z, X_{(-n)})$ obtained after marginalizing out $\boldsymbol{\theta}$ is a student-t distribution (Gelman et al., 2004, p.88). Similarly, the probability $P(z_n = i|Z_{(-n)})$ after marginalizing out the stick variables is equal to the expected value $\mathbb{E}[\pi_i|Z_{(-n)}]$ and can be found by applying Eqn. (6), where $a_i^* = a_i + N_i^{(-n)}$, $b_i^* = b_i + \sum_{j=i+1}^{\infty} N_j^{(-n)}$ and $N_i^{(-n)}$ is the number of data associated with cluster $i$ if we remove $x_n$ from the pool. A similar modification to Eqn. (7) (substituting $N_i^{(-n)}$ for $N_i$, etc.) gives the prediction rule for the simplified parameterization $\{\gamma_i\}$.

Finally, we need to compute a normalization constant $\lambda(n)$ by summing the (infinite number of) terms over the cluster index $i$ in Eqn. (9). Fortunately, the predictive distributions $P(\mathbf{x}_n|Z)$ for empty clusters are identical, allowing them to be lumped together. Thus if $K$ is the index of the maximum occupied cluster, $\lambda(n)$ can be computed by summing over only $K+1$ terms.

Gibbs sampling proceeds by drawing a uniform random variable $u$ in the range $[0, 1]$, and determining the first cluster label $i^*$ for which $\sum_{i=1}^{i^*} P(z_n = i|Z_{(-n)}, X)$ exceeds $u$. We then set $z_n = i^*$, and iterate. Note that, by finding each of these probabilities sequentially, one avoids computing more than the finite number ($i^*$) of required values.

## 3.3 MIXING OVER CLUSTERS

Empirically, this sampler will get stuck in a local minimum where a cluster is always associated with the same label. This state of affairs has no real impact on the DP sampler discussed in Section 2, because a relabeling of the clusters does not change the probability of the data, and sampling is performed over equivalence classes where the actual label does not matter. However, in the stick–breaking formulation the label does matter, because each label has a different prior probability over its size. Hence, swapping the labels between two clusters $i$ and $j$ *does* result in a different probability.

The solution is to make sure the sampler mixes effectively over cluster labels. To achieve that we have introduced mixing moves over cluster labels. There are many possible options for mixing moves, and without exhaustively searching over possibilities we used the following method which worked well for us in practice[3]. Between each sample of an assignment variable, with probability equal to the static parameter $M_i$, mixing move $i$ was proposed. $M$ was set based on how fast we observed the sampler converging over cluster labels. If the problem had many local minimum and the sampler was converging slowly, as in the DDP experiment described in Section 6, $M$ was set to a high probability, such as 1.

The first move "*label-swap*" randomly chooses two cluster labels according to their prior probabilities $p_i = \gamma_i/\gamma$, then proposes to swap them and accepts this move by using the usual Metropolis-Hastings accept/reject rule. Because the proposal distribution is symmetric and the predictive distributions $P(\mathbf{x}_n|Z = i, X_{(-n)})$ are unchanged under this swap, the acceptance rule only needs to consider the change in the probability of cluster assignment variables $P(Z_{\texttt{new}})/P(Z_{\texttt{old}})$. Since the distribution on cluster sizes depends on the cluster label in the stick-breaking construction, $P(Z)$ will change if the number of data points assigned to a cluster changes. However, due to exchange-ability, we only need to consider swapped clusters. Using Eqn. (7), if clusters $i, j$ are swapped, then after some algebra the acceptance rule is reduced to

$$P_{\text{accept}} = \min\left[1, \frac{\prod_{n=0}^{N_j-1}(\gamma_i + n) \prod_{n=0}^{N_i-1}(\gamma_j + n)}{\prod_{n=0}^{N_i-1}(\gamma_i + n) \prod_{n=0}^{N_j-1}(\gamma_j + n)}\right]$$

where $N_i$ is the number of data points in cluster $i$ before the swap. Our second move, "*label-permute*", randomly permutes all the cluster labels with an index smaller than or equal to some index sampled from the prior $p_i = \gamma_i/\gamma$. The Metropolis-Hastings rule is the same as for "*label-swap*", except now the range of cluster labels that were permuted must be considered.

---
[3]One reviewer suggested another move operator which we believe would work well but did not have time to test: select a label according to its prior probability and propose to swap it with the $k^{th}$ next largest label.

Empirically, it was observed that the "*label-permute*" move improves convergence most early in the chain, when many labels may be disordered. On the other hand, "*label-swap*" is most beneficial when the clusters labels have mostly converged, but a few remain disordered. As we will see in the experiments, these moves suffice to mix over cluster labels.

## 4 ILLUSTRATIVE EXAMPLE

To illustrate the effect of poor mixing between cluster labels, we generated the symmetric dataset in Figure 1(a). We use a standard normal-Wishart prior centered at zero and tuned so that most of the time two clusters best explain the data. The central data-case should be assigned with probability 0.5 to the left or right cluster. To test this we used a prior on cluster sizes with $a_i = 5$ and $b_i = 0.1$. We then run the sampler for 5000 iterations (discarding the first 100 for burn-in) with and without the extra mixing moves. We measure the average association between the central data-case and all other data-cases, where two data-cases are associated if they have the same label. The results are shown in Figures 1(b) and 1(c). Clearly, without mixing the prior favors association with one cluster (right block of 25 data-cases) over the other, but due to symmetry this should not be the case. The extra moves clearly undo this effect.

## 5 DEPENDENT DIRICHLET PROCESSES

There is a very natural extension of the infinite mixture models described in Section 3.2 to multiple, coupled models called a dependent Dirichlet process (DDP) (MacEachern, 2000). We consider a finite number of such models, indexed by $t$ (representing time in our application), and couple the models at the level of the parameters $\boldsymbol{\theta}$. In our setup, at each time point we have a number of observations that we wish to cluster. However, there is no correspondence between these observations; for instance there may be very many observations at one time slice and very few (or none at all) at another. The model draws statistical strength from neighboring time slices by imposing smoothness between the means and possibly the covariances and cluster sizes of neighboring clusters. As an example we will look at radar images of weather patterns, where the color index determines the number of fictitious observations at a certain spatial location. We begin by describing the model and Gibbs sampler as proposed in MacEachern (2000), then identify some problems with this sampler and address them by introducing additional moves.

Let us consider a family of $T$ joint distributions as in Section 3.2, one for each time slice. Each of these $T$ distributions is coupled through a joint prior distribution on the parameters $P(\boldsymbol{\theta})$,

$$P(X, Z, V, \boldsymbol{\theta}) = P(\boldsymbol{\theta}) \prod_t P(X_t|Z_t, \boldsymbol{\theta}_t) P(Z_t|V_t) P(V_t)$$

where each individual term $P(X_t|Z_t, \boldsymbol{\theta}_t)$, $P(Z_t|V_t)$, and $P(V_t)$ are modeled as in Eqns. (8a)–(8c) respectively. Note that one could also use a joint distribution over stick-lengths $V$, but for the sake of simplicity we will consider them to be independent. This allows us to integrate out $V$, resulting in a prediction rule[4] $P(z_{n_t} = i|Z_{-n_t})$ similar to the expected value in Eqn. (7) but replacing $N$ by $N_t$ and $N_i^{(-n)}$ by $N_{t,i}^{-n_t}$. We also assume the prior weights are the same at all time slices.

The Gibbs sampler alternates sampling assignment variables from $P(z_{n_t}|X_t, Z_{-n_t}, \boldsymbol{\theta}_t)$ and parameters from $P(\boldsymbol{\theta}_t|\boldsymbol{\theta}_{-t}, X_t, Z_t)$. The equation for the first conditional is similar to Eqn. (9), but without integrating out $\boldsymbol{\theta}$,

$$P(z_{n_t} = i|X_t, Z_{-n_t}, \boldsymbol{\theta}_t) \propto \qquad (10)$$
$$\mathcal{N}[\mathbf{x}_{n_t}; \boldsymbol{\mu}_{z_{n_t}}, \Sigma_{z_{n_t}}] \, P(z_{n_t} = i|Z_{-n_t}) \quad i \leq K_{\max}$$

where $K_{\max} = \max_t \max_{n_t} z_{n_t}$ is the largest occupied cluster label over all time slices. As in Section 3.2, to sample from Eqn. (10) we must compute a data-dependent normalization term. We can again lump together all clusters which are unoccupied in *any* time slice, i.e., clusters $i > K_{\max}$, ensuring that the required summation is again finite. Note, however, that empty clusters with labels smaller than $K_{\max}$ are sampled and not marginalized out. If analytic marginalization is not possible for $i > K_{\max}$, an alternative sampling-based method is described in Papaspiliopoulos and Roberts (2005).

The distributions at each time slice are coupled through the parameters' conditional distribution,

$$P(\boldsymbol{\theta}_{i,t}|\boldsymbol{\theta}_{i,-t}, X_t, Z_t) \propto \qquad (11)$$
$$\prod_{n_t|z_{n_t}=i} \mathcal{N}[\mathbf{x}_{n_t}; \boldsymbol{\mu}_{z_{n_t}}, \Sigma_{z_{n_t}}] \, \mathcal{W}(\Sigma_{i,t}^{-1}) \, \mathcal{N}(\boldsymbol{\mu}_{i,t}|\boldsymbol{\mu}_{i,-t})$$

---

[4] We have slightly abused notation to avoid additional clutter: for instance when we write $\mathbf{x}_{n_t}$, we mean $\mathbf{x}_{n_t}^t$, i.e. the coordinates at time t of the n'th particle.

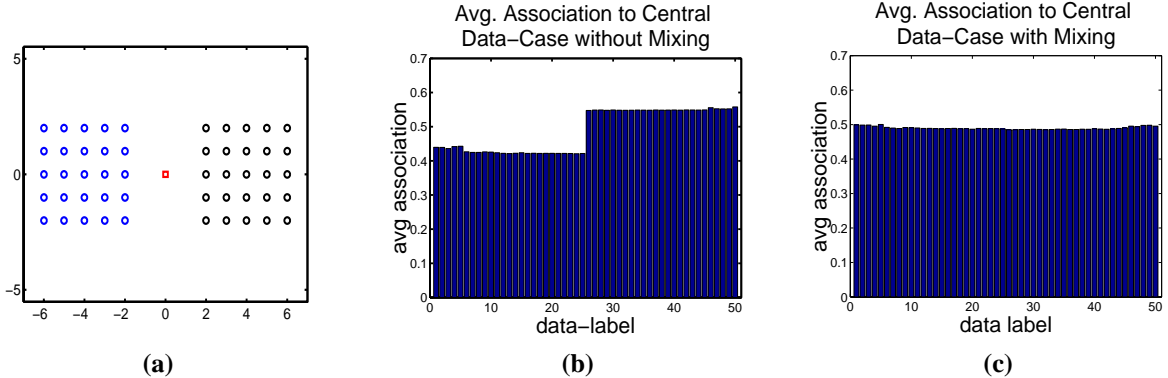

Figure 1: (a) The symmetric data-set to illustrate clustering bias in the absence of mixing moves. (b) Average association of center point to all other data-cases for a sampler without mixing moves. Note the cluster bias. (c) Average association of center point to all other data-cases for a sampler with mixing moves.

Many interesting choices for the joint distribution over parameters are conceivable. For instance, a state–space dynamical model can be chosen for the cluster means (and computation made efficient via a Kalman filter). In our experiments we model the cluster covariances using independent inverse Wishart distributions, while smoothness over the cluster means is enforced through a joint Gaussian distribution $P(\boldsymbol{\mu}) = \mathcal{N}[\boldsymbol{\mu}; \mathbf{m}, \Gamma]$ with $\mathbf{m}$ set to the sample mean of the data and $\Gamma$ given by,

$$\Gamma_{t,t'} = a\exp(-\beta ||t-t'||^\delta) I_d; \qquad \Gamma_t = b I_d \qquad (12)$$

with $I_d$ the identity matrix in $d$ dimensions[5].

The poor mixing over labels reported in Sections 3.3 and 4 turns out to be even more problematic for the DDP. The reason is that the DPs at each time slices are often initialized to have different labels for corresponding clusters. If the DPs were uncoupled the samplers may converge, but when coupled these configurations represent local modes in the probability distribution from which the naive Gibbs sampler has difficulty escaping (see Figure 3). However, the problem can be fixed by introducing additional moves, which explicitly propose label re-orderings and accept or reject them with the standard Metropolis-Hastings rule. In the coupled case we need to propose moves that will improve mixing over labels within each DP, but we also need to propose moves which will bring separate DPs into alignment with each other. The basic extra move we use is a variation on "*label-swap*", where the labels of two clusters $i$ and $j$ are swapped for all time slices in an interval $t \in [t_1, t_2]$. The boundaries are picked in three ways, 1) $t_1 = t_2$, 2) uniformly at random between $[1, T]$ or 3) $t_1 = 1$ and $t_2 = T$ ($T$ is the last time slice). The Metropolis-Hastings acceptance rule is similar to the non coupled case, but it now must consider the change in the coupled variables $P(\boldsymbol{\mu}_{\text{new}})/P(\boldsymbol{\mu}_{\text{old}})$ in addition to the uncoupled assignment variables $P(Z_{\text{new}})/P(Z_{\text{old}})$. These moves were found to improve the mixing behavior considerably.

---

[5]This choice slightly complicates computing the normalization constant because the prior $P(\boldsymbol{\theta}_t) = \mathcal{W}(\Sigma_t^{-1}) \mathcal{N}(\boldsymbol{\mu}_t)$ in the predictive distribution $P(\mathbf{x}_n|Z) = \int \mathrm{d}\boldsymbol{\theta} P(\mathbf{x}_{n_t}|\boldsymbol{\theta}_t) P(\boldsymbol{\theta}_t)$ is no longer conjugate to the likelihood term. We resolve this issue by pre-computing this quantity through Monte Carlo integration.

Additionally, in the stick-breaking construction the expected prior probabilities for a cluster ($\gamma_i/\gamma$) are size ordered. This induces dependencies between the sizes of clusters across all times slices that may be uncalled for or at least hard to control. In Griffin and Steel (2006) the ordering in which the sticks are combined into the prior probabilities are themselves subject to a random process, resulting in some level of control of these dependencies. We view this as an inherent shortcoming of the DDP in the stick-breaking representation that needs further investigation.

## 6 MODELING STORMS

For illustrative purposes we chose to apply Gibbs sampling over coupled infinite mixture models in the stick breaking representation to the problem of tracking storms. Specifically, ten doppler radar images separated by 30 minutes each were converted to data which was then clustered as described in Section 5. Results were compared between DDP with and without "*label-swap*" mixing moves. The results demonstrate that DDP without mixing moves do not mix well over labels, and consequently have trouble clustering across time slices. Note that "*label-permute*" is not used because it results in swap proposals that are rarely accepted for the storm data set.

The experimental data consists of 896 data points, split between ten time slices. The data points were created by using pixel location, normalized to $[0, 1]$, to determine the coordinates, and pixel intensity to determine the number of data points generated per pixel. The parameters for the DDP described in Section 5 were set as follows in all experiments. In Eqn. (12) $a$ and $\delta$ are set to 1 and $\beta$ to .005. The scale matrix for the inverse Wishart was set to $.01 I_d$

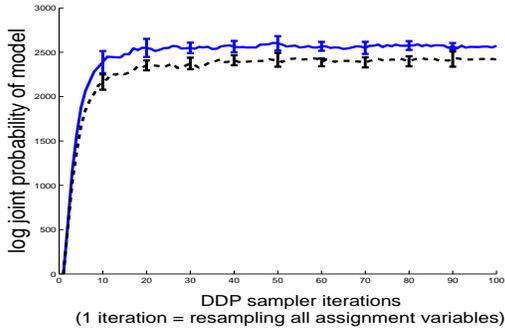

Figure 2: Log joint probability for the DDP sampler with extra mixing moves (solid) and without extra mixing moves (dashed). Curve averaged over 5 runs. The DDP with mixing moves ends up in a region of higher probability.

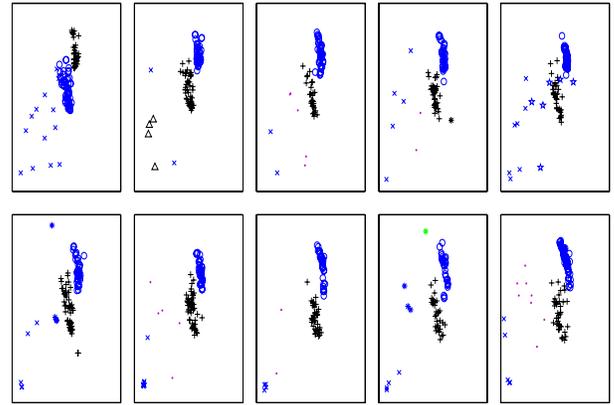

Figure 4: Same as in figure 3 but for DDP sampler with mixing moves. Almost all clusters have been brought into correspondence.

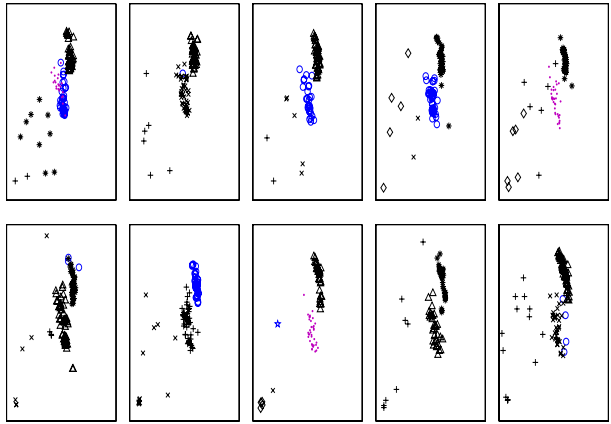

Figure 3: Cluster assignment with highest joint probability for DDP sampler without mixing moves. Points with different markers have different cluster labels. Clusters are not in correspondence.

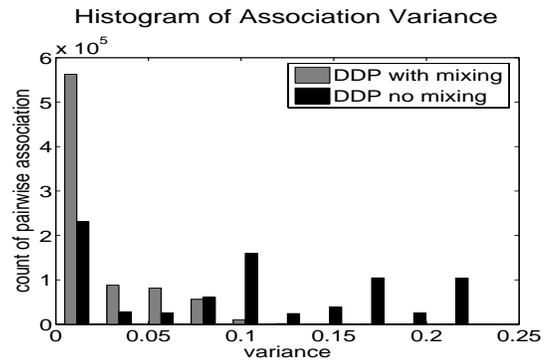

Figure 5: Histogram of the variance of association matrices computed across different sampling runs. The smaller variance of the DDP with mixing moves indicates that it consistently finds a good clustering and does not get stuck in local modes.

with $d + 1$ degrees of freedom.

Because our goal was to find the distribution over cluster assignments, we would like to compare convergence of Gibbs samplers with and without mixing moves in terms of the cluster assignment variables. However, because the mixing moves frequently reassign cluster labels it can appear that the Gibbs sampler with swapping converges more quickly than it actually does. To more accurately gauge mixing behavior, we examine three quantities. First, we compare the log probability of the samplers with and without mixing moves. Second, we qualitatively compare the clustering results. Finally, we compare the variance over the *association* between each pair of points over several Gibbs runs; as we discuss, this measures the convergence of the cluster assignment variables while being invariant to the actual cluster labeling.

First, Figure 2 illustrates how the DDP without mixing moves gets stuck in local modes of the joint distribution. Both Gibbs samplers move quickly toward regions with higher probability than their initial, random starting point, but at the end of the experiment, the Gibbs sampler without additional mixing moves has failed to find the same (much more probable) mode(s) as the Gibbs sampler with additional moves, as indicated by the gap between their respective joint log-probabilities.

Next, Figure 3 shows the cluster assignment with highest joint probability found by DDP without mixing. Clusters appear reasonable within each time slice but lack coherency across time slices. This is in contrast to Figure 4 for DPP with mixing, in which associated clusters have been brought into correspondence (except for the first time slice). In each experiment, Gibbs sampling was performed for 100 iterations, half of which were used for burn-in and half for estimating posterior quantities. Each iteration of the Gibbs sampling consisted of sampling all assignment variables, interleaved with sampling class parameters, as described in Section 5.

Because cluster labels are unidentifiable, we can not use them to evaluate clustering performance between Gibbs sampling with and without mixing moves. Instead, we use an association matrix to infer the clustering behavior. An association matrix is a $N \times N$ matrix in which each element $(i, j)$ is set to 1 if the assignment variables $x_i$ and $x_j$ are equal (i.e. $z_i = z_j$), where $N$ is the total number of data points. This representation is clearly invariant to a re-ordering of the cluster labels. By averaging the association matrices we observe during Gibbs sampling (again, over 50 samples), we can estimate the posterior probability that $x_i$ and $x_j$ came from the same cluster.

To assess convergence of each sampling procedure, we use the (element-wise) variance of the mean association matrix over several (10) runs of Gibbs sampling. If the sampling method is mixing rapidly, these runs should each be consistent with one another (low variance), as they approach the true probabilities of association. If, however, the sampling mixes slowly, they are more likely to be inconsistent (high variance). Figure 5 shows a histogram of the element-wise variances observed across runs for each method. The additional moves show much lower variance, indicating faster mixing and a more consistent clustering.

## 7 CONCLUSION

In this paper we have explored Gibbs sampling of (potentially coupled) infinite mixture models in the Stick-breaking representation. This study is important because the stick-breaking representation allows more flexible modeling of one's prior beliefs, for instance the distribution over cluster sizes. The DDP formalism in particular, where models can be coupled at the level of their parameters, is very general and we expect many future applications that require efficient Gibbs sampling.

Our main finding is that the Gibbs samplers proposed in various manuscripts suffer from poor mixing over cluster labels. Although the impact of this observation is mild on a single infinite mixture model, its effect in the DDP is much more pronounced because labels of corresponding clusters at different values of the covariate (e.g. time) can be mixed up. We show that with the inclusion of a few extra moves in the sampler this problem can be resolved. A disadvantage of the DDP formalism seems to be that dependencies are introduced by the fact that the prior weights on clusters are size ordered. These dependencies do not decay with distance, and seem difficult to control (see however Griffin and Steel (2006)).

**Acknowledgments**

This material is based upon work supported by the National Science Foundation under grants No. IIS-0535278, IIS-0431085, ATM-0530926, and SCI-0225642. The authors would like to thank both Professor H. Ishwaran and the anonymous reviewers for their feedback and insights.